\documentclass{aa}  

\usepackage{ulem}
\usepackage{graphicx}
\usepackage{color}
\usepackage{txfonts}
\usepackage{subcaption}         
                              
\usepackage{lscape}            
                                
\usepackage{placeins}

\begin{document}

   \title{The hybrid Nova Vul 2024 (=V615 Vul)}
   
   \titlerunning{The hybrid Nova Vul 2024 (=V615 Vul)}

   \subtitle{ }

   \author{P. Valisa\inst{1}
	  \and
	  U. Munari\inst{2}
	  \and
	  I. Albanese\inst{3}
	   }
           
   \authorrunning{P. Valisa et al.}

   \institute{
             ANS Collaboration, c/o Astronomical Observatory, 36012 Asiago (VI), Italy\\ \email{paolo.valisa@astrogeo.va.it}
	     \and
             INAF National Institute of Astrophysics, Astronomical Observatory of Padova, 36012 Asiago (VI), Italy
	     \and
             Dept. of Physics and Astronomy, University of Padova, Asiago
	     Astrophysical Observatory, 36012 Asiago (VI), Italy
             }

   \date{Received October 20, 2025 / Accepted February 16, 2026}

 
  \abstract
  %
  %
  {Among galactic novae, many remain poorly characterized due to
   heterogeneous data, observational challenges, or heavy reddening along
   the Galactic plane.  A detailed characterization is particularly crucial
   for novae exhibiting peculiar features, such as those belonging to the
   hybrid class.}
  %
  %
  {We present the spectroscopic and photometric evolution of the heavily
  reddened Nova Vul 2024 (=V615 Vul) from discovery on 29 July 2024
  to well into its nebular phase.}
  %
  %
  {We obtained daily optical, absolute-fluxed spectroscopy in both low- and
   high-resolution echelle modes, providing a detailed account of the
   spectral evolution.  Photometric data were primarily sourced from 
   AAVSO, while refined astrometry was performed to assess the positional
   coincidence with a potential progenitor.  Robust determinations of
   $t_3$ and $E_{B-V}$ were achieved, allowing for the application of
   the Maximum Magnitude–Rate of Decline (MMRD) relation; the resulting
   distance is in excellent agreement with 3D Galactic extinction maps.}
  %
  %
  {Nova Vul 2024 is a very fast nova ($t_3$=10.7±0.5 d) located at a
  distance of 5.0±1.0 kpc, suffering from a large reddening
  ($E_{B-V}$ =1.6±0.1).  Around maximum light, it exhibited a Fe II-type
  spectrum with very broad emission lines (FWZI$\approx$5800 km\,s$^{-1}$) and
  high-velocity P-Cygni absorptions.  Following $t_3$, coinciding with
  the emergence of hard X-ray emission, the nova displayed pronounced
  photometric oscillations primarily driven by continuum variations. 
  Simultaneously, He/N features developed alongside Fe II lines, classifying
  V615 Vul as a rare hybrid nova.  During the nebular phase, the ionization
  level increased, reaching [Fe VII] and likely [Fe X].  The ejecta showed
  no neon overabundance and expanded ballistically, as indicated by the
  constant widths, profiles, and castellation of the high-resolution
  emission lines.}
   { }


   \maketitle

   \nolinenumbers

\section{Introduction}

Classical novae are powered by a sudden thermonuclear runaway of hydrogen-rich 
material that is accreted onto the surface of a white dwarf from a low-mass companion 
in a close binary system \citep{1995cvs..book.....W}. 
The outburst results in a dramatic increase in luminosity and the ejection of 
non-equilibrium CNO-burning products at velocities ranging from a few hundred 
to a few thousand km/s.
Based on early-time ejecta spectra, novae are classified by 
\citet{1992AJ....104..725W} into two main spectroscopic classes: those 
characterized by prominent Fe II emission lines and those dominated by He/N 
emission lines. A small fraction of novae exhibit hybrid behaviour, evolving 
from the Fe II to the He/N class.

Nova Vul 2024 (= PNV J19430751+2100204 = V615 Vul; NVul24 for short) was
discovered on 2024 July 29.832 UT, just a few hours before passing through
photometric maximum, in wide field unfiltered images obtained by
\citet{2024CBET.5423....1S} with a 135mm f/2.0 telephoto
lens in the framework of the New Milky Way Survey operating at the
Astroverty astrofarm in Nizhnii Arkhyz, Karachay-Cherkessia, Russia.

The object was classified soon thereafter as a reddened classical nova by
\citet{2024ATel16743....1T} based on an optical spectrum obtained on July
30.47 UT.  Compared to Taguchi’s spectrum, all emission lines appeared
weaker in an echelle spectrum recorded by \citet{2024ATel16746....1V} nine
hours later on 30.87 UT, a decline typical of novae quickly rising towards maximum
brightness.  On Jul 30.87 UT the Swift satellite pointed to the nova, but
no X-ray emission was detected \citep{2024ATel16751....1S}.  Detection of
hard X-rays from shock-heated plasma within the nova ejecta finally came
around day +11 and persisted through to day +24 \citep{2024ATel16788....1S}.

In this paper we present a detailed analysis of NVul24 based on the
photometric evolution as recorded by the American Association of Variable 
Star Observers (AAVSO) and the Zwicky Transient Facility (ZTF) 
\citep{2019PASP..131a8003M} and on our intensive spectroscopic monitoring 
that collected 31 spectra of the nova, on a daily basis during the first 20 
days following its discovery, and then at a reduced cadence up to day +118, 
well into the advanced nebular phase.  
Particular attention is given to the transition from the initial FeII-type 
to a He/N-type spectrum on day +11, which identifies NVul24 as a new member 
of the elusive class of hybrid novae.  
The emergence of the He/N spectrum occurred simultaneously
with the onset of photometric oscillations and with the transformation of
the H$\alpha$ profile from that typical of optically thick conditions to
that characteristic of an optically thin shell.

\section{Observations}

\subsection{Spectroscopy}

Spectra of NVul24 were recorded both at high resolution and at
low dispersion with long-slit spectrographs.  A logbook of the
spectroscopic observations is provided in Appendix A.  We collected a total
of 31 spectra distributed over 28 individual nights spanning the period from
discovery to day +118 after the outburst, when NVul24 had already faded
seven magnitudes from maximum and was well into the nebular phase.

Most of the high-resolution (Echelle) spectra have been obtained with the
Varese Schiaparelli Observatory 0.84m telescope, which is operated by ANS
Collaboration.  The telescope is equipped with an Astrolight Inst.  mk.III
Multi-Mode carbon-fibre spectrograph, which in the Echelle configuration
covers the 4250-8900 \AA\ range \citep[cf.][for an optical description and
performance evaluation of these multi-mode
spectrographs]{2014CoSka..43..174M}.  Cross-dispersion is achieved with a
N-SF11 prism, and the detector is an SBIG ST10XME CCD camera
(2192$\times$1472 array, 6.8$\mu$m pixel, KAF-3200ME chip with micro-lenses
to boost the quantum efficiency).  The spectral resolution obtained with a
R2 79 l/mm grating is 20,000 for a 1-arcsec slit and CCD binning=1$\times$,
lowering to 15,000 for a 2-arcsec slit, and 12,000 for CCD
binning=2$\times$.  Only the two reddest orders are affected by inter-order
gaps, between 8239-8243~\AA\ and 8554-8574~\AA.

One additional Echelle spectrum, at a later epoch when NVul24 became fainter
than R=13, was collected using the 1.82m telescope + REOSC Echelle
spectrograph which is operated in Asiago by INAF (Italian National Institute
of Astrophysics).  The 3500-7350~\AA\ interval is covered over 32 orders
without inter-order gaps by an Andor DW436-BV camera (housing a 2048x2048
array, 13.5 $\mu$m pixel size, E2V CCD42-40 AIMO model CCD).  The resolving
power is 22,000 for the standard 1.8-arcsec slit width.

Low-dispersion spectra were acquired with the B\&C spectrograph on the 1.22m
telescope operated in Asiago by the University of Padova.  The spectra were
obtained with a 300 ln/mm grating blazed at 5000 \AA, which covers the
3200-8000~\AA\ range at 2.3 \AA/pix dispersion.  The detector was a 2048$\times$512 
ANDOR iDus DU440A E2V 42-10 back-illuminated CCD.  This setup
offers excellent sensitivity in the near-UV down to 3200~\AA.

All observations at all telescopes were conducted with the slit oriented
along the parallactic angle, for optimal sky-subtraction and
flux calibration.  Data reduction for all telescopes has been performed in
IRAF and has included all usual steps for bias, dark, flat, long-slit sky
subtraction, wavelength calibration, heliocentric correction, and flux
calibration. The absolute flux calibration was obtained
via nightly observations of the spectrophotometric standard HR 7596, 
located on the sky close to the nova.   
This allowed us to join all Echelle orders into a single 1D-fluxed spectrum covering
the whole recorded wavelength range.  The flux zero point of each
spectrum has been checked against the near-simultaneous $B$$V$$R$$I$ 
data from AAVSO.

\subsection{Photometry}

BVRI photometric observations of NVul24 were retrieved from
the American Association of Variable Star Observers (AAVSO) database.  The
high cadence of these data up to day +50 allowed for binning into 0.5-day
intervals, effectively reducing the scatter.  Within each bin, the average
difference between the minimum and maximum V-band magnitude is 0.4 mag, with
a standard error of the mean of approximately 0.1 mag.  At later epochs, the
lower sampling density increases the uncertainty per interval.

The discrepancies between measurements from different
observers likely arise from a lack of transformation to the standard BVRI
system.  Furthermore, the intense H$\alpha$ emission, which overlaps with
the red wing of the V-band transmission profile, may significantly affect
the photometry depending on the specific filter response of each observer. 
Finally, ZTF g- and r-band photometry was obtained through the ZTF Forced
Photometry Service \citep{2019PASP..131a8003M}.

\section{Photometric evolution}

\begin{figure}[ht!]
   \centering
   \includegraphics[width=\hsize]{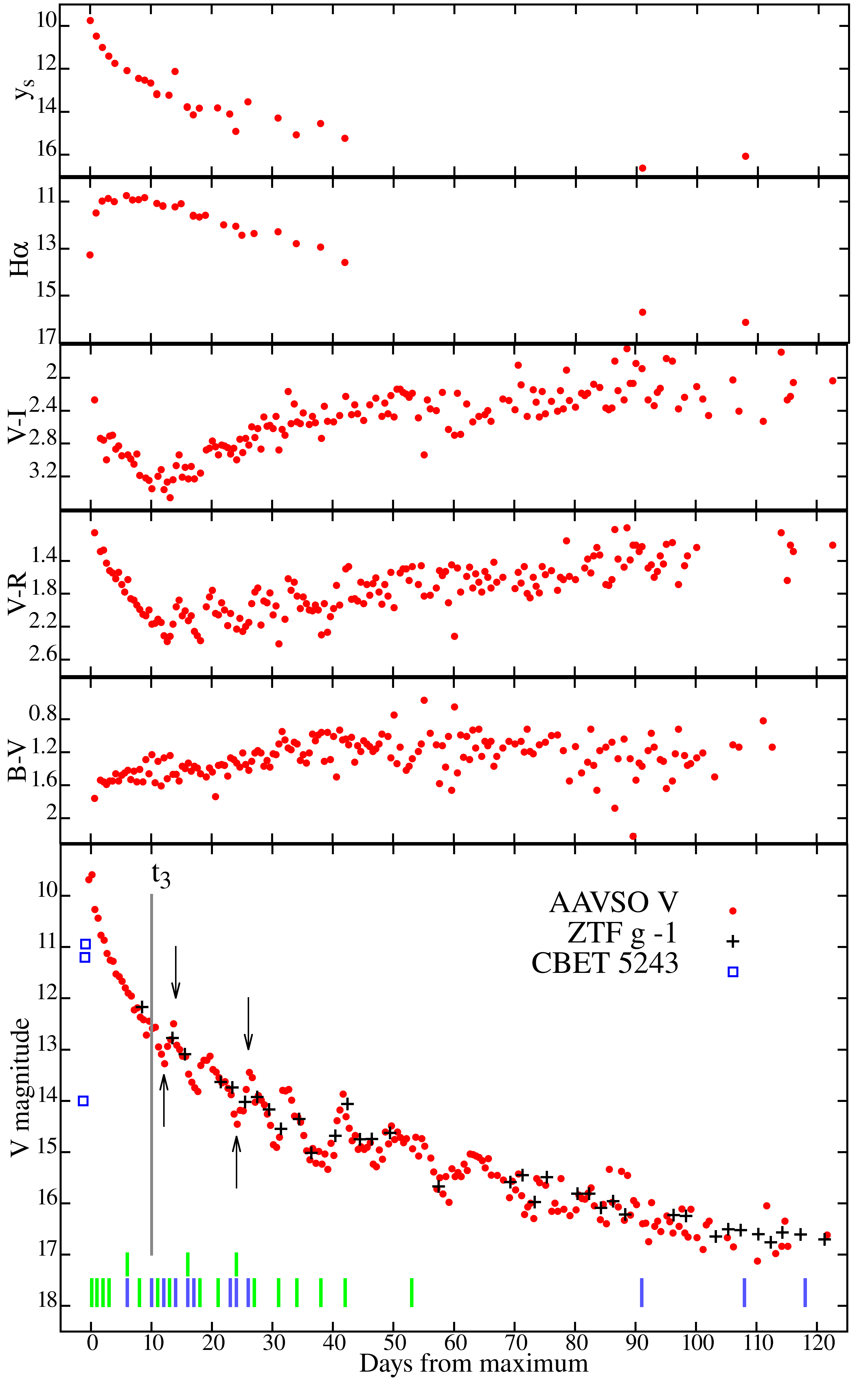}
      \caption{Colour and light curves of Nova Vul 2024.  The Str\"omgren 
        $y$-band magnitude and the H$\alpha$ emission-line flux,
        expressed in magnitudes, are obtained by integrating the flux-calibrated
        spectra. Black dots are ZTF $g$ magnitudes rescaled to AAVSO $V$
        magnitudes by adding a $-$1 mag offset.  The observing epochs are
        counted as days passed since optical maximum (2024 Jul 30.8 UT). 
        The green ticks in the $V$-band panel mark the epoch of Echelle
        observations, while the blue ticks are the epochs of low resolution
        spectra.  The arrows point to epochs of spectra shown in
        Fig.~\ref{oscillations} that were obtained at minima and maxima
        during oscillation period.}
         \label{phot1}
   \end{figure}

The AAVSO and ZTF light- and colour-curves of NVul24 covering the first 120
days of the outburst are presented in Fig.~\ref{phot1}.  The rise to
maximum, the passage at peak brightness, and the early decline were rapid,
smooth, and well-mapped by the observations.  From the light curve in
Fig.~\ref{phot1} we derive as the time of maximum in $V$ band HJD=2460522.3
(= 2024 Jul 30.8 UT), which we will adopt as the $t_0$ reference epoch in
this paper.

The rise time to maximum brightness was less than 2 days, since the latest
reported non-detection was 2024 Jul 28.840 UT (at a limiting mag 14.5, cf.  CBET
5423).  The time required to decline by two and three magnitudes from
maximum were $t_2$=5$\pm$0.5 and $t_3$=10.7$\pm$0.5 days, respectively,
which are typical of very fast novae according to the classification scheme
of \citet{1995cvs..book.....W}.  The small value of $t_2$ makes NVul24
one of the fastest FeII novae of recent years.  Other very fast novae of the
FeII class include N Cyg 2005 \citep[V2361 Cyg, $ t_2=6.0 $
days;][]{2007ApJ...662..552H}, N Aql 1999 \citep[V1494 Aql, $ t_2=6.6 $
days;][]{2000A&A...355L...9K}, N Cyg 2007 \citep[V2467 Cyg, $ t_2=7 $
days;][]{2011ApJS..197...31S}, and N Cyg 2008 \citep[V2468 Cyg, $ t_2=7.8 $
days;][]{2011A&A...526A..73I}.  Recent novae faster than NVul 24 all belong
to the He/N class, such as N Cyg 2001 N2 \citep[V2275 Cyg, $ t_2=2.9 $
days;][]{2002A&A...384..982K} and N Her 2021 \citep[V1674 Her, $ t_2=1 $
days;][]{2021ApJ...922L..10W}.

  \bigskip

Overall, the light curve of NVul24 resembles the 'O' class (oscillating
novae) according to \citet{2010AJ....140...34S}, with an initially smooth 
decline and quasi-periodic oscillations of $\approx$1 magnitude amplitude, 
starting around the passage at $t_3$ and lasting to about day +70 (the 
increasing noise in the AAVSO data at later epochs makes it difficult to 
discern if oscillations were persisting after that epoch).   The time-scale of
the oscillations increased from an initial value of 5-6 days, to 9-12 days
around day +70.  Some other novae share this same photometric behaviour
(of a lengthening with time of the oscillation time scale),
including V603 Aql, V1494 Aql \citep{2011A&A...526A..73I}, V4745 Sgr
\citep{2005A&A...429..599C}, V888 Cen \citep{2010AJ....140...34S} and V2467 Cyg 
\citep{2009AN....330...77P}.

\begin{figure}[ht!]
   \centering
   \includegraphics[width=\hsize]{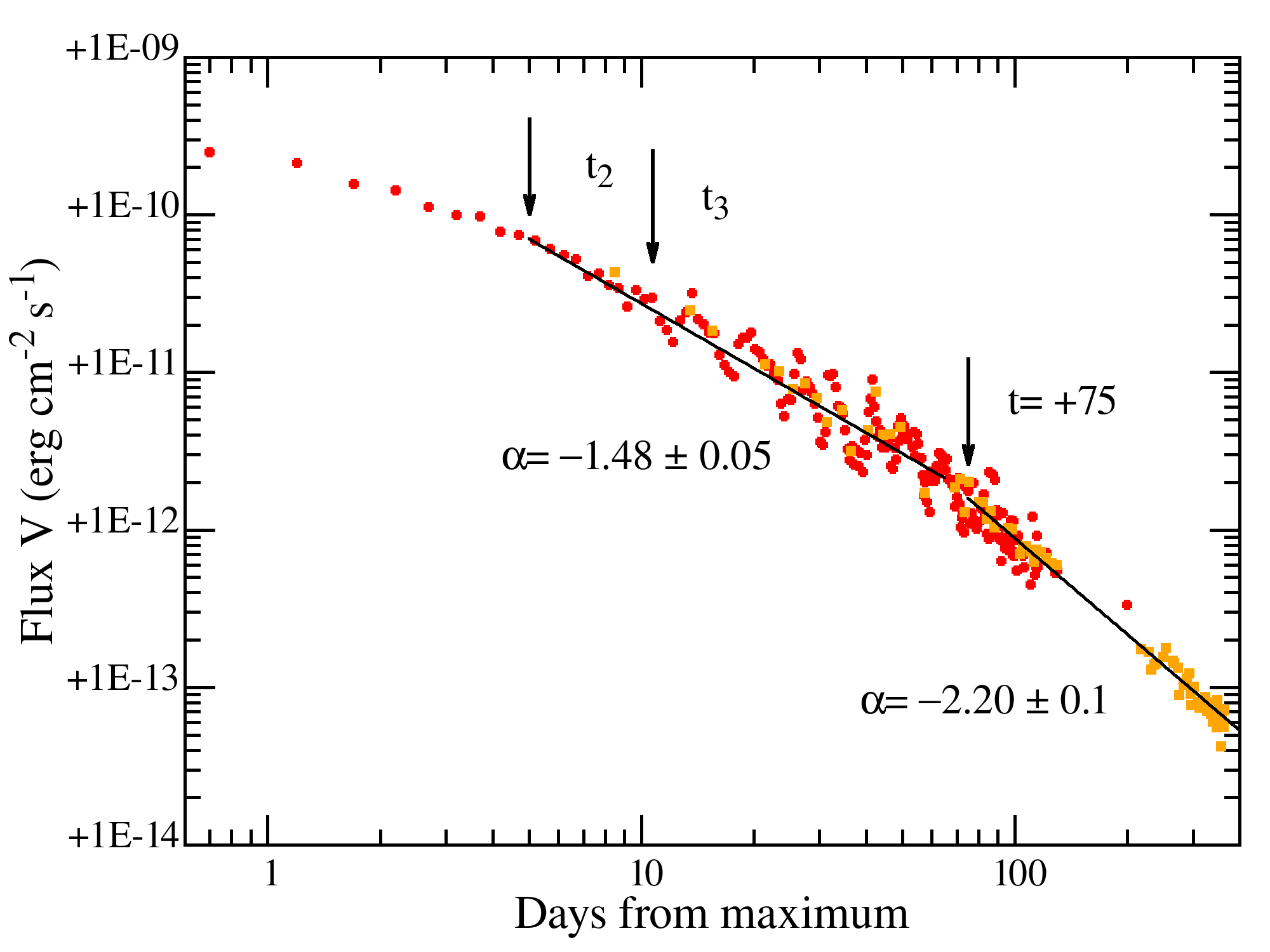}
      \caption{Log-log plot of the time evolution of the flux radiated by NVul24
      through the $V$ band.  The fit with a broken power-law 
      of the form $F^{V} \propto (t-t_0)^{\alpha} $ is overplotted (lines in
      black) and the corresponding values of $\alpha$ are quoted.
      Red dots are from AAVSO magnitudes, binned each half-day. 
      Orange dots are from ZTF $g$ magnitudes rescaled to $V$ magnitudes by 
      adding a $-$1~mag offset.}
         \label{phot2}
   \end{figure}

In general, O-class light curves after the peak are well described by a
broken power-law of the form $F^{V}_{\rm decline} \propto (t-t_0)^{\alpha}$
\citep{2006ApJS..167...59H, 2010AJ....140...34S}.  As shown in
Fig.~\ref{phot2}, NVul24 conforms to that with $\alpha=-1.48 \pm 0.05$ from $t_2$ to
about day +75, and $\alpha=-2.2 \pm 0.1$ after that.  The change in the slope of the
broken power-law may relate to the end of the nuclear burning in the shell
of the WD, implying a cooling of its surface temperature and consequently a
decline in the feeding to the ejecta of high-energy photons able to contrast
the ion recombination. 
Although information on the SSS phase in NVul24 is not available, evidence 
suggesting that the end of the SSS phase may influence the decay slope of several 
emission lines was reported by \citet{2023A&A...674A.139A} in a detailed study 
of RS Oph.
\citet{2006ApJS..167...59H} preferred a different
interpretation, with the transition in the broken power-law marking the time
at which stops the wind that they propose is continuously blowing-off the WD
since early times in the outburst.

  \bigskip

The colour evolution of NVul24 follows the average evolution observed in
novae, as discussed by \citet{2025MNRAS.538.2339C}.  The $(B-V)$ evolution
towards bluer colours reflects the shift at shorter wavelength of the maximum
of the blackbody-like emission of the pseudo-photosphere while it retraces
through the ejecta as their density decline because of the expansion.  To
separate the role of emission lines and underlying continuum in governing
the behaviour of colours, on our accurately fluxed spectra we have measured
separately the flux of the H$\alpha$ emission line and that in the y Stromgren
band (which is remarkably free for the presence of significant emission
lines through the whole evolution of a normal nova \citep{2008ASPC..401..206H} 
and plotted them at the top of Fig.~\ref{phot1}.  
The flux zero-point for the $R$ band has been adopted for H$\alpha$ too.  
It is rather clear how the $(V-R)$ colour is dominated up to $t_3$ by the great 
contribution from H$\alpha$, similarly to $(V-I)$ colour by OI 8446 which is 
pumped by Lyman-$\beta$ fluorescence. 
Past $t_3$, both H$\alpha$ and OI 8446 accelerate their decline matching the
pace of the underlying continuum, with the net result that both $(V-R)$ and
$(V-I)$ became bluer with time for the same reasons as for $(B-V)$.

\section{Reddening and distance estimates}

An estimate of the reddening of NVul24 was obtained by
\citet{2024ATel16746....1V} from an echelle spectrum recorded on July 30.869
UT.  A value $E_{B-V}$=1.7 mag was derived from the equivalent width (EW)
the diffuse interstellar band (DIB) at 6614 \AA\ , using the calibration by
\citet{2014ASPC..490..183M}, and a similar $E_{B-V}$=1.75 mag was derived
from the EW of interstellar KI 7699 \AA\ line, following the calibration of
\citet{1997A&A...318..269M}.

A high reddening affecting NVul24 is confirmed by other indicators.  For the
DIBs at 5780 and 5797~\AA\ we measure an EW of 0.78 and 0.34~\AA,
respectively, implying $E_{B-V}$=1.6 for both of them from the calibrations of
\citet{2013ApJ...774...72K}.  \citet{1987A&AS...70..125V} reported the
intrinsic colour of novae as $(B-V)_\circ$=+0.23 and $-$0.02 when passing at
maximum brightness and at $t_2$, respectively.  Comparing with the AAVSO
lightcurve in Fig.~\ref{phot1}, we derive $E_{B-V}$=1.5 on both such epochs. 
Taking the unweighted average of all above estimates, the reddening
affecting NVul24 is $E_{B-V}$=1.6$\pm$0.1, which will be adopted in this
paper.

The MMRD relation (magnitude at maximum vs. the rate of decline) is frequently
used to estimate the distance to a nova.  Its most recent formulation by
\citet{2019A&A...622A.186S} is based on Gaia DR2 parallaxes and makes use of
$t_3$ in the $V$ band. 
The MMRD relation applied to $t_3$=10.7$\pm$0.5 days for NVul24 
returns an estimated absolute magnitudes $M_V$=$-$8.9$\pm$0.4, that coupled 
with $E_{B-V}$=1.6$\pm$0.1, leads to an estimate of 5$\pm$1 kpc as the distance 
to the nova. 
A similar distance is derived from the brightness after 15 days ($M_{v15}$),
that was proposed by \citet{1955Obs....75..170B} to be the same for all
novae.  The calibration by \citet{2019A&A...622A.186S} provides
$M_{v15}$=$-$5.71$\pm$0.4 mag, that when applied to NVul24 returns the same
5$\pm$1 kpc distance. Such a distance places NVul24 at $120\pm$20pc below
the Galactic plane and within the Orion-Cygnus spiral arm.

It should be noted, however, that significant scatter remains in the MMRD $M_V$ 
and $M_{v15}$ of single novae as already pointed out by \citet{2000AJ....120.2007D} 
and also evident in recent surveys on extragalactic novae in M31, M77, M87 
(\citet{Shara_2017}, \citet{2018MNRAS.474.1746S}). 
It remains unclear whether this scatter arises from intrinsic differences among 
novae or from observational uncertainties.
The main sources of observational uncertainty arise from the determination of 
reddening along the line of sight, from the accurate measurement of the time 
of maximum and $t_3$ which can be affected by undersampling, irregularities, 
and oscillations in the light curve.

The 5~kpc distance to NVul24 inferred from the MMRD relation is well supported 
by the reddening progression along the line of
sight to the nova, as mapped by the IPHAS \citep{2014MNRAS.443.2907S} and
Bayestar19 \citep{2019ApJ...887...93G} extinction maps.  These maps clearly
indicate that a reddening of $E_{B-V}$$\approx$1.6 is reached only at distances
greater than 4~kpc.

\section{Astrometry and identification of the progenitor}\label{sec:astrometry}

The coordinates of NVul24, derived when the nova was close to its peak
brightness, have been reported in CBET 5423 as R.A.  = 19:43:07.51, Decl. 
=+21:00:20.4 (J2000.0).  In order to check and improve on them, we observed
NVul24 on 2024 October 22 when the nova had declined to $V\approx$16 mag,
making the astrometric observations against field stars more robust.  To
this aim we used the Ferrante 0.36m telescope operated by Societ\'a
Astronomica Schiaparelli in Hakos Farm (Namibia) and the Varese 0.84m
telescope.

\begin{figure}[ht!]
   \centering
   \includegraphics[width=\hsize]{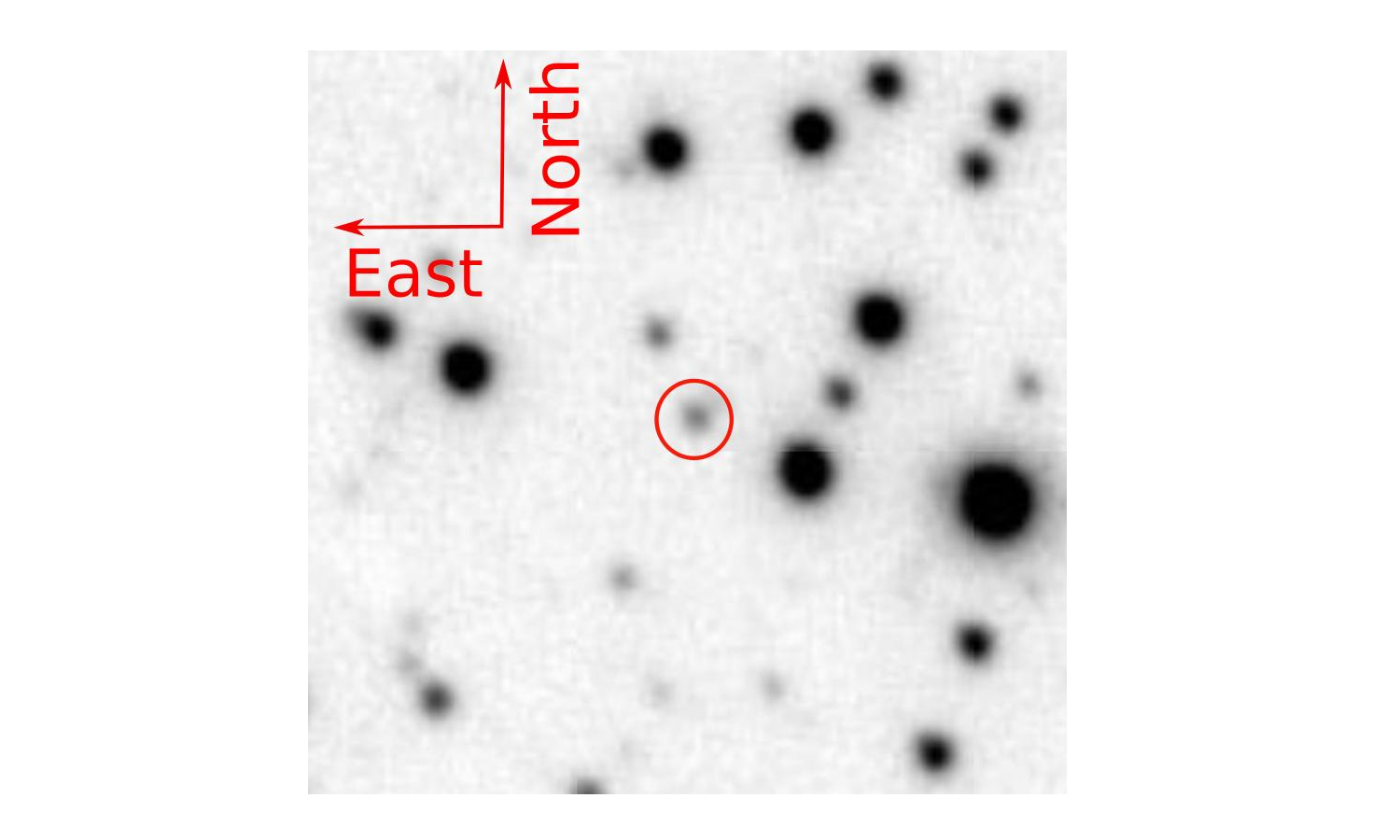}
      \caption{PanSTARRS $r$ image cropped around the position of NVul24. 
	The size is 30x30 arcsec.  The red circle has a radius of 1.5
	arcsec, and it is centred at the accurate position we measured for
	NVul24 (cf.  sect.~\ref{sec:astrometry}).  The star close to the
	centre of the circle is GAIA DR3 1825912166611947136 (of $G$=19.82
	mag), which possibility to be the progenitor of the nova is
	discussed in sect.~\ref{sec:astrometry}.}
         \label{astrometry_1}
   \end{figure}

Our improved astrometric position for the nova, using the Gaia DR3 catalogue 
as reference, is R.A.=19:43:07.498 and Decl.=+21:00:21.23 
(0.075 arcsec radius error), and the 1.5 arcsec radius red circle plotted 
in Fig.~\ref{astrometry_1} is centred on it. 
A faint star reported both in Gaia DR3 and PanSTARRS DR2 lies within 0.12 
arcsec of our position for the nova.

There are pros and cons about considering it the progenitor of NVul24.
Let's begin with the problematic aspects.

Gaia DR3 names the star 1825912166611947136 and lists for it $G$=19.82,
$B_P$=20.82, and $R_P$=18.46 mag, a slow-moving proper motion, and a parallax
$\pi$=1.53$\pm$0.40 mas \citep{2016A&A...595A...1G,
2023A&A...674A...1G}.  Taken at face value, such parallax would suggest a
distance $\leq$1~kpc, much closer than the 5~kpc above estimated for the
nova.  The associated error (27\%) is however too large for an entirely safe
inversion of the parallax to derive an accurate value for the distance; to
achieve this, \citet{2015PASP..127..994B} and \citet{2018A&A...616A...9L}
recommended an error on the parallax that should not exceed 20\%.

According to \citet[][its Fig.~5.4]{1995cvs..book.....W}, there is a good 
correlation between $t_2$ and the outburst amplitude for a nova, with faster
novae being those showing the largest amplitude. At $G$=19.82 mag, the candidate
progenitor using the geometric parallax distance would imply an outburst amplitude 
of just $\sim$11 mag, when the minimum amplitude according to Warner's relation 
(corresponding to the accretion disk of the progenitor being seen face-on) would 
be 13 mag.

On the pro side, we note that the Bayesian posterior photogeometric distance 
estimate from \citet{BailerJones2021}, which incorporates colour and magnitude 
information in addition to the parallax, would give a much larger distance 3.6~kpc, 
with a 68\% confidence interval of 1528--5097~pc.

In addition, PanSTARRS DR2 lists for the star in the circle $g'$=21.50,
$r'$=20.04, $i'$=19.07, $z'$=18.55, and $Y$=18.19 mag.  Correcting them for
the reddening affecting the nova ($E_{B-V}$=1.6 mag) returns an intrinsic
rather blue colour $(g-r)_\circ$=$-$0.1, the same value reported by
\citet{2023MNRAS.524.4867I} as the mean for cataclysmic variables observed
by the SLOAN survey.  A similar conclusion would be reached by considering
Gaia $B_P$ and $R_P$ data.

Moreover, the field around the nova is sparsely populated as illustrated
in Fig.~\ref{astrometry_1}, and the probability of a chance superposition
within $\sim$0.1 arcsec of a field star with the real progenitor seems
improbable, especially considering the rarity of intrinsically blue
stars among field objects.

We considered the possibility to invest a sizeable amount of observing time
in recording a classification spectrum for the faint star in the circle of
Fig.~\ref{astrometry_1}, to confirm or disprove it being the progenitor of
the nova. We however restrained from that considering that - only
one yr past the eruption - the nova ejecta could have not yet diluted
sufficiently into the interstellar medium to avoid interfering in the
process, especially if the presence of a H$\alpha$ in emission may be looked
for as evidence of accretion being back at work.  Waiting a few more years 
seemed a safer choice.

A fainter Gaia source with $G=21.01$ mag is located at a distance of 1.1
arcsec from our position of the nova, well outside the astrometric ellipse
error.  Parallax and proper motion are not available for this star in Gaia
DR3.

There is no 2MASS source at the position of NVul24.  Around the nova
position, the detection of infrared sources appear complete to $J$=16.8,
$H$=15.0, and $K_s$=14.3 mag.  Adopting colour-dependent reddening relations
for the 2MASS system from \citet{2003A&A...401..781F}, at the distance and
reddening above estimated for NVul24, this is equivalent to say that the
absolute magnitude of the progenitor is $M(K_s)$$\geq$0.1.  Comparing with
the distribution in $M(K_s)$ of known novae as derived by
\citet{2025CoSka..55c..47M}, this limit is sufficient to exclude the
presence of a cool giant or a bright sub-giant as the donor, but leaves
ample margins to accomodate a fainter sub-giant.  Such a midly evolved
companion to the WD, by its intrinsic larger luminosity, would naturally
account for the reduced outburst amplitude observed in NVul24.

Summarizing, the main (or better the only) problem with firmly identifying
Gaia 182591216661194713 as the progenitor of NVul24 is the apparently too
large Gaia parallax, albeit affected by a significant error, while the
intrinsic colour and the spatial coincidence argue strongly in favour, and
the oserved outburst amplitude of 11 instead of a minimum of 13 mag can be
circumvented by a slightly evolved donor star which presence is compatible
with the upper limit to its brightness in the near-IR.

As a final comment, we have accessed the Harvard historical photographic
plate collection digitized as part of the DASCH programme
\citep{2012IAUS..285...29G}, but found no past detection of NVul24, implying
that ($i$) no previous nova outburst has been observed, and ($ii$) if the
progenitor underwent dwarf-nova outbursts, these were either rather
unfrequent or characterized by a low amplitude in brightness (a few
magnitudes at most, so that the progenitor remained below the detection
threshold of Harvard photographic plates).

\section{Spectral evolution}

\begin{figure*}[ht!]
   \centering
    \includegraphics[width=17cm]{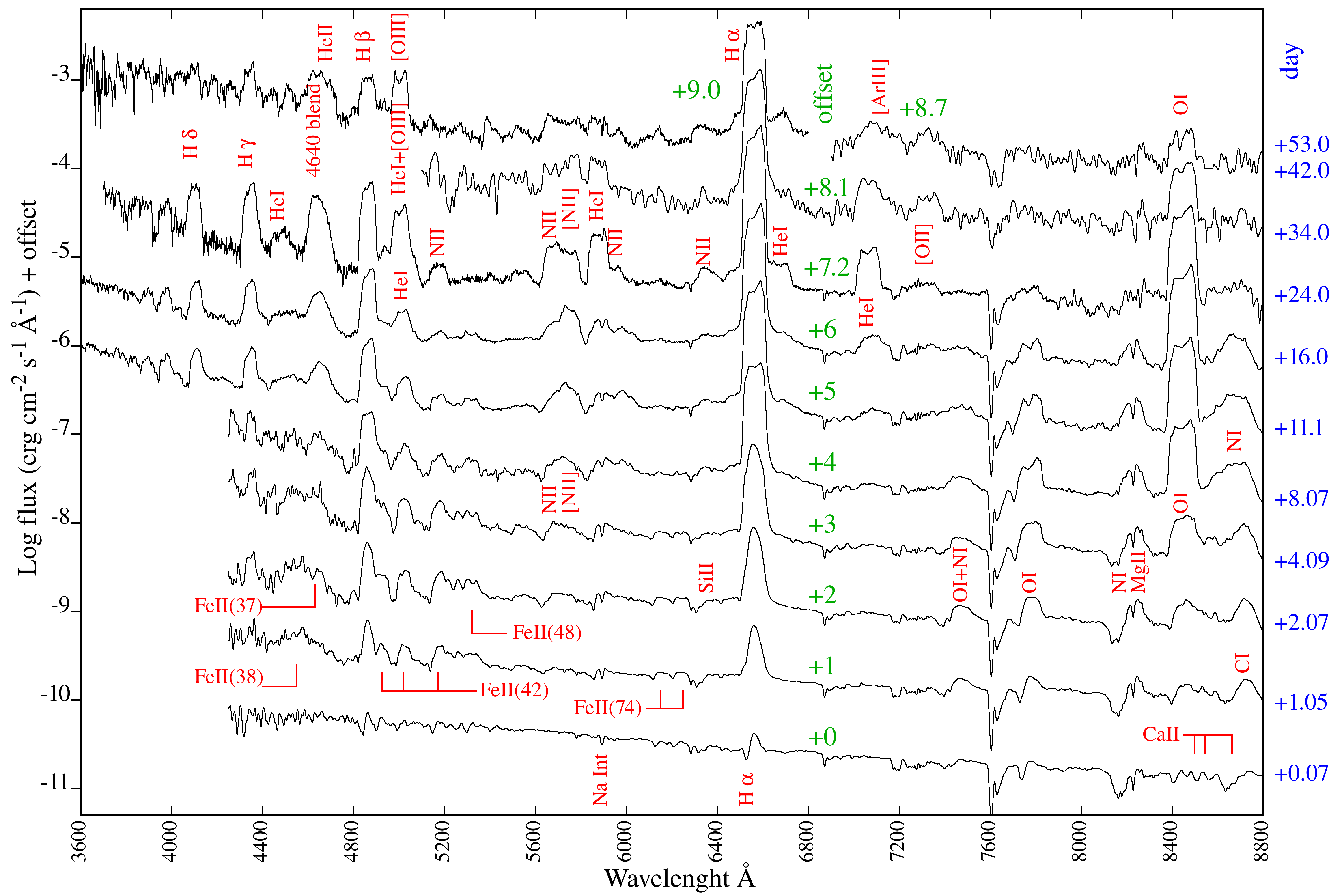}
      \caption{Spectral evolution of NVul24 for the first 50 days of the
        outburst.  The sequence of sample spectra were obtained in low
        dispersion with the Asiago 1.22m+B\&C (day +11, +16, +24) and in
        high resolution with the Varese 0.84m+ Echelle (day +0, +1, +2, +4,
        +7, +34, +42).  The latter were Gaussian smoothed and binned to
        reduce their spectral resolution to match that of the Asiago
        1.22m+B\&C spectra, for an easier intercomparison.  The native
        absolute fluxing of all the spectra has been checked against the
        AAVSO photometry of Fig.~\ref{phot1}, and dereddened by
        $E_{B-V}$=1.6.  The offset in flux applied to the spectra for a
        clearer plotting is indicated in green.}
         \label{spectral_evolution}
\end{figure*}

The initial spectral evolution of Nova Vul 2024 up to day +53 is presented
in Fig.~\ref{spectral_evolution}, by combining a sample of spectra selected
from those we have recorded (listed in Table~\ref{table:3}).

\subsection{Maximum, early decline, and absorption systems}

We obtained an echelle spectrum for each of the first five days, during
which the changes were most pronounced for this very fast nova.  The
spectrum of July 30 corresponds to the photometric maximum (day +0.07),
dominated by the continuum of the fireball phase.  The only lines to appear
faintly in emission are H$\alpha$, H$\beta$, OI 7772 and FeII multiplet 42,
all accompanied by absorptions on the blue side.  The absorption to
H$\alpha$ has a bulk velocity of $-$1450 km/s and extends to $-$2270 km/s,
values that for OI 7772 turn to $-$1310 km/s and $-$2000 km/s, respectively. 
Blue-shifted absorptions flank also OI 8446, SiII 6347/71, CaII 8252 and
the CaII triplet 8498, 8542, 8662 \AA.

One day past maximum (our spectrum for day +1.05, cf.  Table~\ref{table:3}),
as typical for very fast novae, the spectra of NVul24 displayed a retracing
continuum and emission lines of increasing flux, very broad and flanked by
weakening P-Cyg absorptions.  The OI 7772, 8446 and the FeII emission lines
(multiplets 37, 38, 42, 48 and 74) gave to NVul24 spectra the aspect typical
of FeII-class novae following the classification by
\citet{1992AJ....104..725W}.  The equivalent width of H$\alpha$ increased
from $-$25~\AA\ on day +0.07 to $-$158~\AA\ on day +1.05 (see also
Fig.~\ref{phot1}), when it displayed a FWHM=2470 km/s.  The broad
blue-shifted absorption to H$\alpha$ became narrower and moved to $-$1720
km/s, with possible secondary components at $-$1250 and $-$2200 km/s.  They
can be identified with the principal absorption system according to the
nomenclature of \citet{1960stat.book..585M}.  At the same time, the broader
diffuse enhanced absorption system also appeared, centred at $-$2800 km/s
and extending up to $-$3500 km/s at about twice the average velocity of the
principal absorption.  A detailed description of the evolution of
absorptions systems for H$\alpha$ for the first three days is provided in
Appendix~\ref{appendix_abs_systems}.

The strong emission seen during the first three days around 8725~\AA\ may be
identified with CI 8727~\AA, following \citet{2002A&A...390..155D}.  Other
strong emission features in the red appears to originate from blends: that
at 8650 should come from multiplets \#6 of SI and \#1 of NI, at
8220~\AA\ between multiplets \#2 of NI and \#7 of MgII, that at 7477 from
multiplets \#3 of NI and \#55 of OI.  In Fig.~\ref{spectral_evolution} the
feature at 8220~\AA\ displays an apparently double-peaked profile, but this
is a spurious effect caused by the superposition of a strong telluric
absorption and an Echelle inter-order gap (8239-8246 \AA).

\subsection{H$\alpha$ and OI 8446 \AA~evolution}

\begin{figure}[ht!]
   \centering
   \includegraphics[width=\hsize]{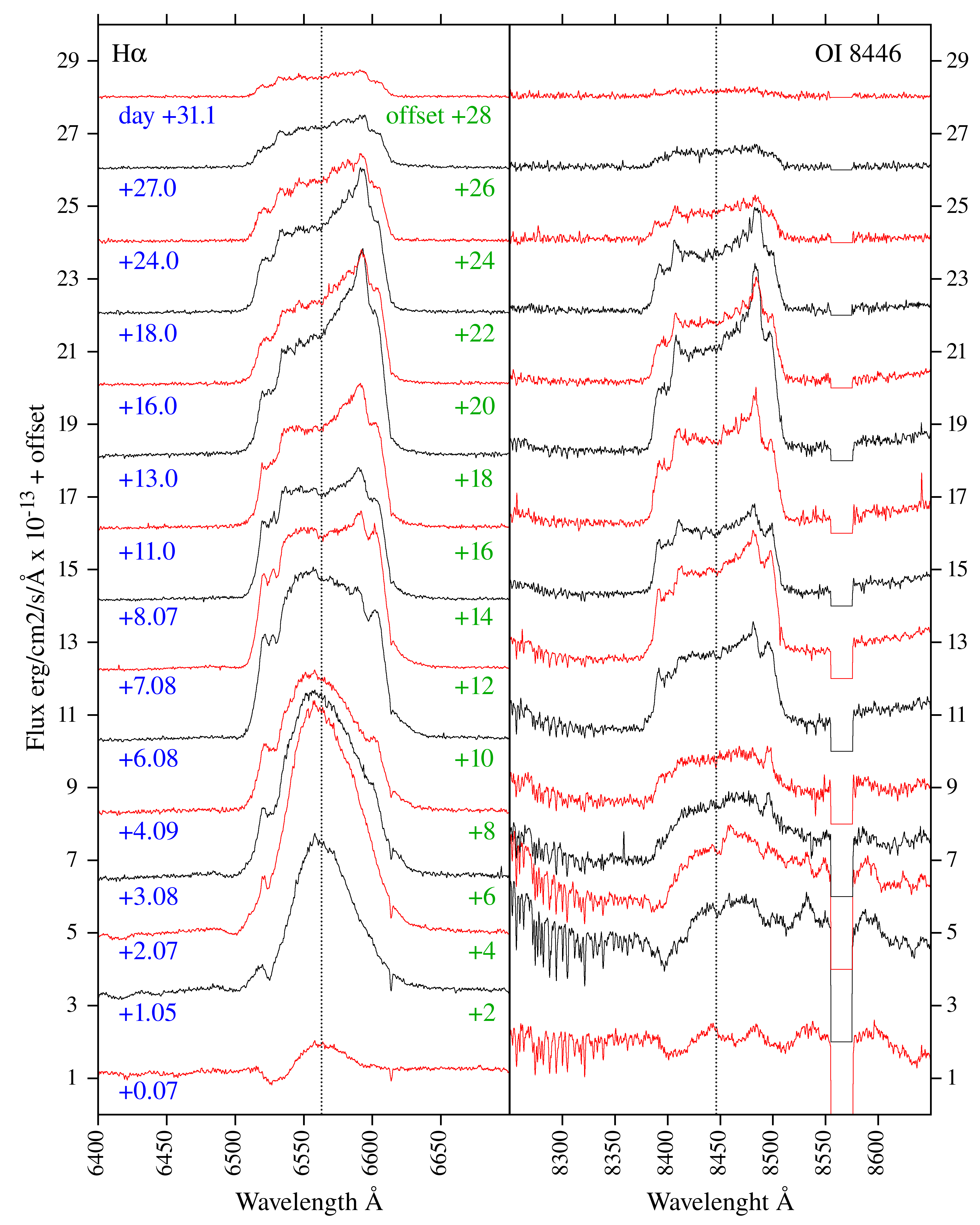}
   \caption{Evolution of H$\alpha$, and OI 8446 \AA\ emission lines.  The
	H$\alpha$ line profile has been corrected for telluric absorbptions. 
	No correction was applied to OI 8446 \AA\ line because telluric
	absorptions ends redward 8377 \AA\ .  The drop in intensity between
	8555 and 8575 \AA\ is artificial and represents an inter-order gap
	in the Echelle spectra.}
   	\label{Halpha_OI}
\end{figure}

The evolution in flux and profile of H$\alpha$ and OI 8446 \AA~line is
presented in parallel in Fig.~\ref{Halpha_OI}.  

Around the passage at $t_2$ on day +5, the shape of H$\alpha$ changed from a
rounded profile to a “wedding cake” arrangement, with a rectangular pedestal
of FWHM=4300 km/s and a superimposed narrower component of FWHM=2750 km/s. 
The boxy pedestal to H$\alpha$ turns visible at the same time the broken
power-law of Fig.~\ref{phot2} enters the $\alpha$=$-$1.48 branch, which
according to \citet{2006ApJS..167...59H} should mark the transition from
optically thick to thin conditions in the wind that they propose blows off
the WD of novae during their early evolution.  The H$\alpha$ pedestal
freezed its profile to FWHM=4300 km/s and FWZI=5800 km/s for the rest of the
nova evolution past $t_2$.  \citet{1992AJ....104..725W} sub-divided the FeII
class of novae based on the value of FWZI, with 5000 km/s separating the FeIIb
(for 'broad') from the FeIIn (for 'narrow') types.  At FWZI=5800 km/s NVul24
belongs to the FeIIb novae, which \citet{1992AJ....104..725W} noted are the 
best candidates to become hybrid novae.

Starting from day +2 and up to day +30, OI 8446 \AA\ became the second
most intense emission line after H$\alpha$, with a similar profile and the
same width.  From day +7 to day +21 the dereddened intensity of OI 8446 \AA\
line remained at around 25\% of H$\alpha$.  The intensity of 8446 \AA\
triplet OI line, under pure recombination and optically thin conditions,
should be 3/5 of the quintuplet OI 7773 \AA\ line, but the dereddened flux ratio
8446/7773 in NVul24 reached $\sim$30 by day +25 and was still $\geq$10 in the
spectrum for day +25 when the 7773 \AA\ line was last visible.

The inversion in intensity between these two OI lines in novae ejecta is
usually attributed to Lyman-$\beta$ fluorescence pumping of the OI 8446 \AA\
line, as first described by \citet{1947PASP...59..196B}.  Lyman-$\beta$
photons can be generated in regions where neutral Oxygen is abundant,
through the conversion of H$\alpha$ photons, provided that the $n$=2 levels
of HI are significantly populated by collisions \citep{1977ApJ...216...23S}. 
Such regions, with locally high opacity in H$\alpha$, can be
identified with dense discrete blobs of gas embedded in the thin wind of the
ejecta.  The Doppler signature of their differing radial velocities is
confirmed by the castellated shape of the H$\alpha$ profile shown in
Fig.~\ref{nebular_ha}.

\subsection{Emergence of He/N spectrum and hybrid classification}

A striking feature of Fig.~\ref{spectral_evolution} is the emergence around
day +7 of the NII+OII+NIII blend at 4640 \AA\ as well as the [NII] 5755 +
NII 5678 and NII 5938 + HeI 5876 blends. They are typical of the He/N class
of novae \citep{1992AJ....104..725W}.

\begin{figure}[ht!]
   \centering
   \includegraphics[width=\hsize]{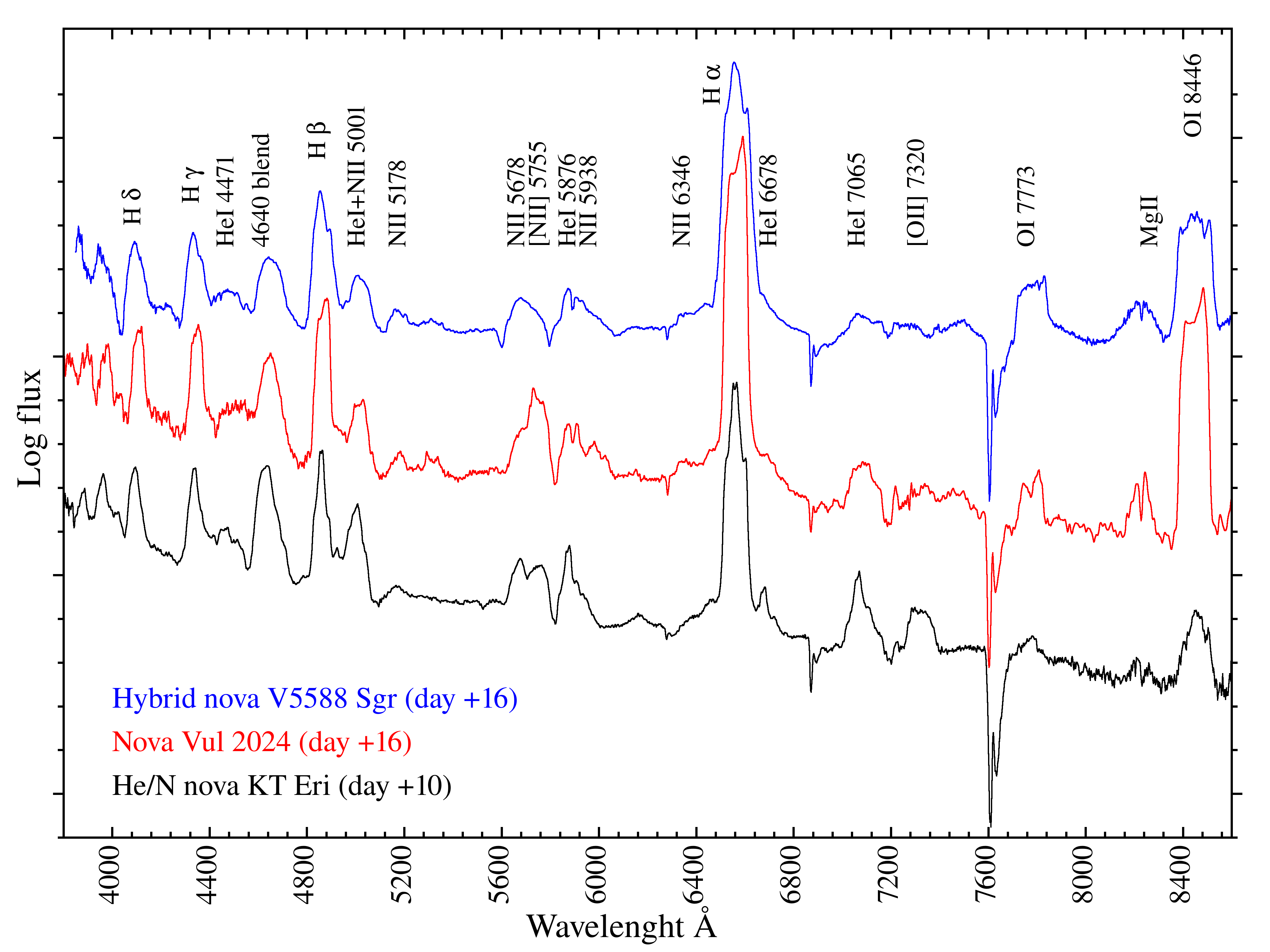}
   \caption{Spectrum of NVul24 for day +16, compared to those of the hybrid 
    nova V5588 Sgr and the He/N nova KT Eri.}
   	\label{confronto}
\end{figure}

As the FeII lines progressively weaken from day +11 onwards, the spectrum of
a He/N-type nova begins to emerge, becoming fully developed between days +16
and +24.  A comparison of the spectrum of NVul24 for day +16 with those of
the hybrid nova V5588 Sgr and the He/N nova KT Eri is presented in
Fig.~\ref{confronto}.  There are only very few novae that have been seen to
evolve from a FeII- to an He/N-type spectrum or to have simultaneously shown
them both.  \citet{1992AJ....104..725W} called them “hybrid” novae and
defined them as “novae that change between the two spectral classes during
the early permitted emission-line phase, evolving from an FeII type spectrum
to that of a He/N type spectrum before forbidden lines appear ”.

The nature of the hybrid spectra observed in some novae remains debated.
\citet{2012AJ....144...98W} suggested that the physical conditions producing He/N spectra 
are consistent with an origin in the white dwarf (WD) ejecta, whereas Fe II 
spectra arise in an extended circumbinary gas envelope, likely originating 
from the secondary star. In hybrid novae, the transition between the two 
spectral classes is attributed to the temporal evolution of the physical 
parameters in these two emitting regions during the post-outburst decline.

More recently, \citet{2024MNRAS.527.9303A} proposed that all novae experience 
an initial He/N phase during the rise to maximum light, followed by an Fe II phase 
and a second He/N phase prior to the onset of the nebular spectrum. 
In this framework, the observed spectral class depends primarily on the 
opacity and ionization state of the ejecta. However, one of these phases 
may frequently remain unobserved because of its short duration.

We present a detailed spectroscopic description of the Fe II–to–He/N transition 
in NVul24, with the aim of contributing to a clearer understanding of the scenario 
underlying its occurrence.
Other recent examples of genuinely hybrid novae are V1674 Her \citep{2021ApJ...922L..10W},
V5588 Sgr \citep{2015MNRAS.447.1661M}, V5114 Sgr
\citep{2012ASInC...6..143A}, and V458 Vul \citep{2008NewA...13..557P}.  A
list complete to the best of our knowledge is provided in
Table~\ref{tabHeN}.  \citet{1991ApJ...376..721W} intially considered the
symbiotic recurrent nova V3890 Sgr as an example of hybrid novae, but as
clearly revealed by the intensive spectral monitoring of its latest
outburst in 2019 \citep[eg.][among many
others]{2020SASS...39..199S,2020MNRAS.499.4814P}, the nova spectrum is of
the He/N type since the beginning, with FeII features arising in the ionized
wind of the late type giant and not in the ejecta,
\citep[see sect.  3.1 of][]{2025CoSka..55c..47M}, and therefore there is 
no relation of V3890 Sgr to real hybrid novae or classical novae at large.

\begin{table*}[ht!]
	\centering
	\footnotesize
	\caption{List of genuine hybrid novae.}
	\begin{tabular}{lcccccl}
	\hline\hline
	 Name         &  Year  & $t_2$         & Oscillations           & [OI]  & Neon overabund. &  Reference  \\
	\hline
    V615 Vul      &  2024  &  5                 & yes               &  no   &           & this paper        \\
	V1674 Her     &  2021  &  1                 & no                &  no   & yes       & \citet{2021ApJ...922L..10W}   \\
	V5588 Sgr     &  2011  &  38                & 6 secondary maxima&  no   &           & \citet{2015MNRAS.447.1661M}     \\  
	V458 Vul      &  2007  &  7                 & yes               &  no   & yes?      & \citet{Poggiani2008V458Vul}  \\ 
	M31N 2006-10b &  2006  & $\approx$ 11       &                   &  no   &           & \citet{Shafter_2011}   \\
	V5114 Sgr     &  2004  &  11                & no                &  yes  & yes       & \citet{2006AandA459875E}  \\
	V444 Sct      &  1991  &  6                 &                   &  no   & yes       & \citet{1994ApJS...90..297W}     \\
	V838 Her      &  1991  &  $\approx$ 4       & no                &  no   & yes       & \citet{1991ApJ...376..721W}   \\
	LMC 1988-2    &  1988  &                    &                   &  no   &           & \citet{1991ApJ...376..721W}  \\
	\hline
	\end{tabular}
    \label{tabHeN}
\end{table*}

\begin{figure}[h!]
	\centering
	\includegraphics[width=\hsize]{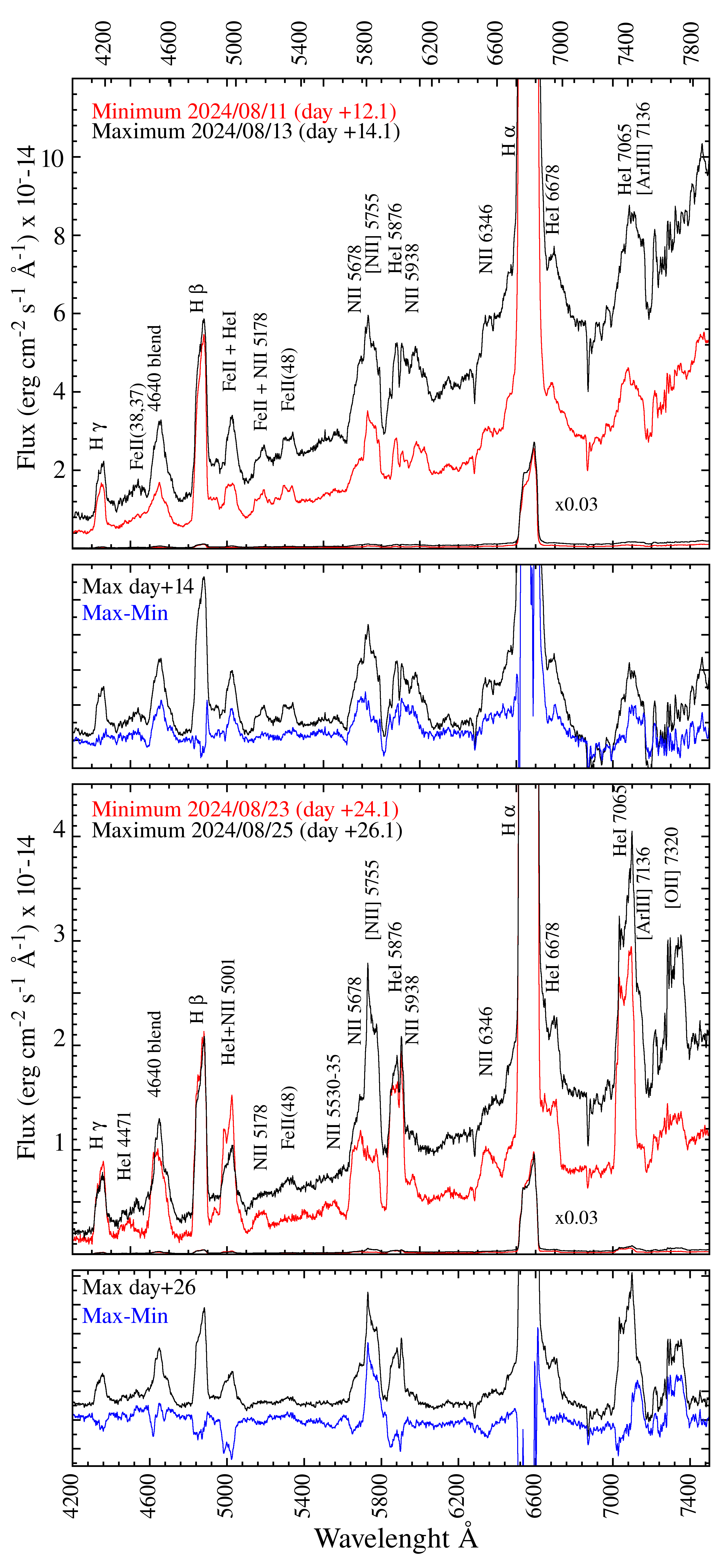}
	\caption{Spectral evolution through oscillations. Large panels: 
    Absolutely fluxed spectral evolution between maxima (black) and 
    minima (red) across the first and third pscillations.  
    To show the evolution of H$\alpha$, the same spectra
	are also superimposed reduced by a factor of 50.  Small panels:
	difference spectrum (blue) between continuum subtracted spectra at
	maximum and at minimum, compared to the spectrum at maximum
	(black).}
	\label{oscillations}
\end{figure}

Due to the large width and consequent strong blend of the emission lines, it 
is not possible to easily isolate the lines associated with the He/N-class spectrum
and compare their Doppler profile with those of the FeII-class spectrum, in
order to derive expansion velocities and the geometry of the ejecta for each
system.  What is certain is that the strong asymmetry affecting H$\alpha$
between days +13 and +18 (cf.  Fig.~\ref{Halpha_OI}) does not replicate in
the profile of NII 5001 \AA.

Novae of the FeIIn type usually display significant [OI] 5755, 6300, and 6364
emission lines, which persist even when the spectrum of the nova turns to
high ionization condition (such as [CaV] or [FeVII] emission lines). 
\citet{1994ApJ...426..279W} evaluated the optical thickness in the [OI]
lines, their collisional excitation, and shielding from the hard ionizing
photons permeating the ejecta, concluding they can only form in isolated,
small and dense blobs/condensations expanding alongside with the much lower
density general ejecta. Interestingly, there are no clearly visible [OI]
lines in the spectra of NVul24, and absence or great weakness of [OI]
lines is the hallmark of hybrid novae (cf. Table~\ref{tabHeN}), reinforcing
the partneship of NVul24 with them.

\subsection{Spectroscopic behaviour during photometric oscillations}

The appearance of the He/N spectrum in NVul24 overlaps with the onset of
photometric oscillations, whose maxima occurred on days +14, +20, +26, +32,
+42, +52, and +63, the latter two values being rather uncertain given the
large noise affecting the AAVSO-based light curve of Fig.~\ref{phot1}.  There
seems to be no direct relation between NVul24 turning hybrid and the
appearance of the oscillations, and no such a correlation may be clearly
established for the other hybrid novae listed in Table~\ref{tabHeN}.

The first (maximum on day +14) and the third (maximum on day +26)
oscillation have been covered by our spectral observations, with spectra
available at maximum and at the preceding minimum (indicated by the
arrows in Fig.~\ref{phot1}).  The spectra at minimum and maximum, and their
subtraction differences, are compared in Fig.~\ref{oscillations}.  In both
cases the increase in nova brightness appears primarily driven by a
continuum turning brighter, in agreement with the fact that the oscillations
are similarly well visible in all the $B$$V$$R$$I$ bands (see
Fig.~\ref{phot4}). From the top two panels of Fig.~\ref{phot1}, it is rather
evident how H$\alpha$ emission passed unscathed through the oscillations,
while the continuum-dominated Stromgren $y$ band show the first oscillations
amplified compared to $B$$V$$R$$I$ bands.

The difference spectrum in Fig.~\ref{oscillations} clearly indicates how the
oscillation on day +14 is accompanied by a flux increase of the emission
lines associated with the He/N spectrum, while little or no variation
observed for FeII and Balmer lines.  A careful inspection of the line shape
in the subtracted spectrum suggests also an asymmetric enhancement for the
HeI lines, with the red part of the profile turning stronger.  The same
behaviour affects H$\alpha$ that simultaneously with the oscillation
developed a peak at +1400 km/s (cf.  Fig.~\ref{Halpha_OI}).  The +1400 km/s
H$\alpha$ peak remained correlated with the photometric behaviour on the
following days: it decreased on day +16 (close to photometric minimum),
strengthened again on day +18 (close to second oscillation maximum of day
+20) and decreased once more by day +24 (near another minimum).

The difference spectrum for the day +26 oscillation (the third one) tells
however the reverse story compared to day +14 oscillation, as it is well
visible in Fig.~\ref{oscillations}.  This time it is the He/N spectrum that
weakens with the rise to peak brightness of the oscillation, expecially the
NII lines, with a parallel large surge in the emission of forbidden lines
such as [NII] 5755, [ArIII] 7136, and [OII] 7325, and the possible reappeance
of FeII again in emission.  Another significant difference compared to the
oscillation on day +14 is the absence of any enhancement of the +1400 km/sec
peak in the H$\alpha$ profile with the passage at maximum brightness.

We recorded Echelle spectra also near the minimum and maximum of the
oscillation of day +42 (the fifth of the series).  By that time the
brightness of the nova had already declined to $V\approx14$, and the
high-resolution Echelle spectra were affected by a low S/N.  Nonetheless,
also for this oscillation an increase in the forbidden lines [NII] 5755,
[ArIII] 7136 and [OII] 7325~\AA\ is observed near maximum brightness.

In light of their spectroscopic evolution, the nature of the photometric
oscillations in NVul24 appear to differ from those sometimes observed in
slow novae, where the spectra around maximum show broadening of the emission
lines — possibly accompanied by high-velocity P Cygni absorptions and lower
ionization conditions — as reported for V1405 Cas
\citep{2023arXiv230204656V}, V5558 Sgr \citep{2008NewA...13..557P} and V4745
Sgr \citep{2005A&A...429..599C}.  Nor can the oscillations in NVul24 be
compared to the flares occasionally observed during the nebular phase, which
typically involve intensity variations in high-ionization lines such as [FeVII]
and [FeX], such as those described by \citet{2022MNRAS.516.4805M} for Nova
Sct 2019 or \citet{DiGiacomo2025nova} for Nova Cas 2021.

It appears relevant to note that the onset of the oscillations of NVul24
coincides with the detection of X-rays with the Swift satellite as reported by
\citet{2024ATel16788....1S}.  The hard spectrum of the X-rays suggest an
origin in shock-heated plasma.  The Swift/XRT count rate peaked on day +16
at 0.068 $\pm$ 0.011 cts/s and declined to 0.042 $\pm$ 0.006 by day +24.
The transition of NVul24 to nebular conditions, with the emergence of [OIII]
lines, also occured during the final stages of the oscillations season.
The potential connection between the observed oscillations
and shock-driven X-ray emission is intriguing; however, the sparse X-ray
data available in the literature for this source limits further discussion
to a speculative level, making a more quantitative analysis currently
unfeasible.

\begin{table}[ht!]
	\centering
	\footnotesize
	\caption{Summary of the main parameters for Nova Vul 2024.}
	\begin{tabular}{rcl}
	\hline\hline
	&& \\
       coordinates&=&19:43:07.498, +21:00:21.23 ($\pm$0.075 arcsec)\\ 
       galactic coord. &=& 57.4084, $-$1.2801  \\
	   $t_{max}$ &=& 2460522.3~{\rm HJD} (2024 ~{\rm July}~ 30.8~{\rm UT})   \\
	   $V_{max}$ &=& 9.6 mag 		     \\
       type &=& initial FeII, then hybrid \\
	   $ t_2$&=&5 $\pm$ 0.5  days    \\
	   $ t_3$&=&10.7 $\pm$ 0.5  days  \\
	   $E_{B-V}$&=&1.6 $\pm$ 0.1  mag  \\
	   $M_V$ &=& $-$8.9 $\pm$ 0.4 mag (from MMRD)\\
	   distance &=&  5 $\pm$ 1  kpc  (from MMRD) \\
           FWZI(H$\alpha$) &=& 5800 km/s\\
           probable &&\\
           progenitor &=& Gaia DR3 1825912166611947136\\
	&& \\
	\hline
	\end{tabular}
	\label{tab_summary}
\end{table}

\subsection{The nebular phase}

In the spectrum of day +53, shown at the top of
Fig.~\ref{spectral_evolution}, the intensity of [OIII] 5007 finally exceeded
that of H$\beta$, marking the onset of the nebular phase, which occured when
the $V$-band luminosity of NVul24 had declined 5 magnitudes below maximum. 
The reduction in density and transition to optically thin condition in the
ejecta led to a significant decrease in the Lyman-$\beta$ fluorescence of
the OI 8446 line, while the increasing ionization favoured HeII over HeI,
with HeII becoming visible on the red side of the 4640 \AA\ blend. 

\begin{figure}[ht!]
	\centering
	\includegraphics[width=\hsize]{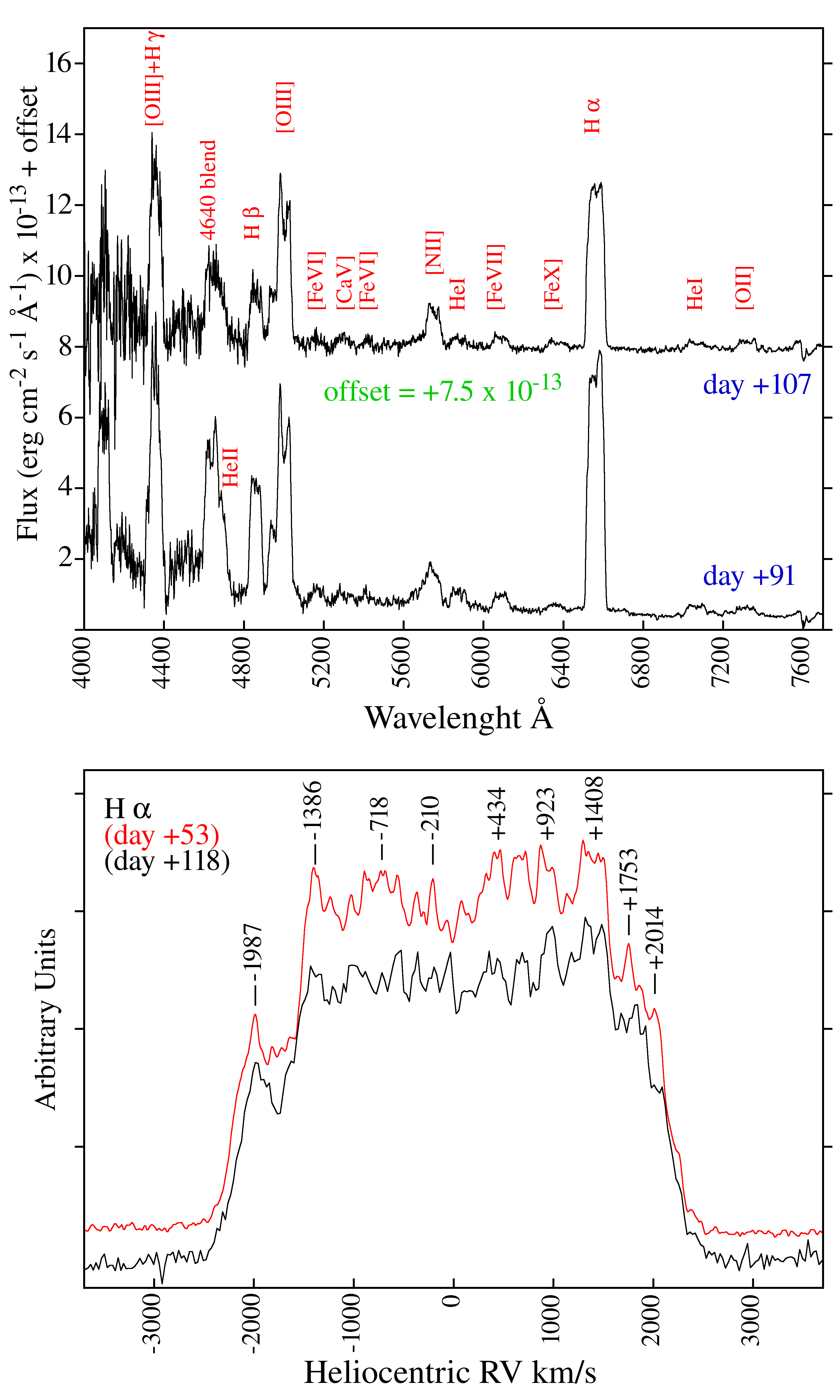}
	\caption{Spectra during nebular phase. Upper panel: reddening-corrected 
        spectra to highlight the evolution of NVul24 during the advanced nebular stage.
	    Lower panel: comparison of the H$\alpha$ profiles for day +53
        and +118, with indication of the velocity (in km/s) for some features.} 
   	\label{nebular_ha}
\end{figure}

Despite the ionization conditions at this stage being favourable for the
excitation of [NeIII], no trace of [NeIII] 3869, 3968 lines was observed in
the spectrum recorded on day +53, or in the subsequent spectra taken on
days +91 and +107 (Fig.  ~\ref{nebular_ha}), when the V-band magnitude 
had already declined to V=16.5 and V=17 respectively.  
The very red colour and overall faintness prevented us from reaching down to
[NeV] 3427. The absence of emission in Ne lines indicates
that NVul24 is not a neon nova.  We have extensively experimented with the
photoionization code Cloudy v23.01 \citep{2023RNAAS...7..246G} for the
physical conditions of NVul24; for the typical overabundance of neon novae,
the [Ne III] 3869 line would be expected to be significantly more intense
than both [O III] 4363, 5007 and H$\beta$, while a Solar abundance agrees
with the lack of detection.

Our last spectrum was secured on day +118, with medium dispersion 
(0.6 \AA\ /pix) allowing a comparison of the H$\alpha$ profile with that 
recorded at the beginning of the nebular phase on day +53 (cf Fig.~\ref{nebular_ha}).  
The two profiles are very similar indicating that the ejecta were expanding 
in a balistic way into the surrounding space; the minor changes visible in the 
peaks of the jagged top, can be ascribed to the low S/N of the later spectrum.

On day +118, the H$\alpha$ profile exhibits a trapezoidal base with FWHM=4340 km/s
(FWZI=5030 km/s) upon which a second boxy and castellated emission component
with FWHM=3130 km/s is superimposed.  Although an analysis of the shape of
the ejecta is beyond the scope of this paper, a similar structure has been
modelled for ejecta composed of equatorial rings and polar caps
\citep{2011MNRAS.412.1701R, 1999MNRAS.307..677G}.  The spectra from days +91
and +107 show the evolution towards lower-density conditions resulting in an
increase in forbidden line intensities relative to permitted one,
alongside a decrease in the Balmer lines, HeI lines and 4640 \AA\ blend.  In
the dereddened spectra from day +107, shown in the Fig. ~\ref{nebular_ha},
the blend of [OIII] 4959+5007 \AA\ reaches the intensity of H$\alpha$ and
[OIII] 4363 \AA\ is perhaps even stronger. The overall spectral evolution
aimed to higher ionization conditions, culminating in the probable
appearance (although at low flux) of [FeX] on the latest spectrum for day
+107.

\section{Conclusions}

We have performed extensive spectroscopic monitoring of
Nova Vul 2024, from its passage at optical maximum to well into the nebular
stage and the appearance of the coronal [FeX] line.  Supported by available
AAVSO and ZTF photometry, we have derived the primary properties of this
very fast and highly reddened nova, which turned out to be a member of the
rare class of hybrid novae.

Both the decline time $t_3$=10.7 days and the $E_{\rm
B-V}$=1.6 reddening were robustly determined; the application of the MMRD
relation to them provided a distance to the nova of $\approx$5~kpc, in fine
agreement with 3D maps of the extinction through the Galaxy.  For the line
of sight to the nova, these maps imply a lower limit of 4~kpc to the
distance.

We obtained refined astrometry of the nova during the
advanced nebular stage, which led to a robust positional match with a G=19.8
mag star present in both the Gaia and PanSTARRS catalogues.  While the Gaia
parallax is inconclusive, the blue colour of this field star, as registered
by both Gaia and PanSTARRS, matches the typical colour of nova progenitors. 
The lack of a 2MASS counterpart excludes a giant or a bright sub-giant as
the donor star in the progenitor, while a faint sub-giant would naturally
explain the somewhat reduced outburst amplitude for NVul24.

The spectrum of the nova around maximum was typical of the
FeII-class, with very broad emission lines (FWZI$\sim$5800) and
high-velocity P-Cyg absorption reaching terminal velocities $\approx$$-$3500
km\,s$^{-1}$.  After $t_3$, the nova began experiencing photometric
oscillations primarily driven by changes in the continuum and the emergence
of hard X-ray emission, while emission lines typical of the He/N-class began
to develop in parallel to those of the FeII-class, characterizing NVul24 as
a member of the rare class of hybrid novae.

During the nebular phase, up to the last spectrum obtained
at day +118, the ejecta remained at relatively high densities; the
ionization kept increasing, passing through [FeVII] and probably up to
[FeX].  The ejecta did not show overabundance in neon, and for most of the
monitored time, they expanded in a simple ballistic pattern, as demonstrated
by the width, profile, and castellation of the high-resolution profiles of
the emission lines.

\section{Data availability}

The spectral data presented in this work were obtained by the authors 
and are available in electronic form at the CDS via anonymous 
ftp to cdsarc.u-strasbg.fr (130.79.128.5) or via 
http://cdsweb.u-strasbg.fr/cgi-bin/qcat?J/A+A/.

Photometric data from AAVSO and ZTF are publicly available and can
be accessed directly through their respective web portals.


\begin{acknowledgements}
We would like to thank the anonymous referee for the
valuable comments that helped to improve the paper and clarify its content.
We also express our gratitude to L. Buzzi for performing accurate astrometry of
NVul24 and A. Milani for downloading hystorical DASCH data.  We acknowledge 
the variable star observations from the AAVSO International Database contributed 
by observers worldwide and used in this research.  This work has been in part 
supported by INAF 2023 MiniGrant Programme (contract C93C23008470001 to UM).  
The ZTF forced-photometry service was funded under the Heising-Simons 
Foundation grant 12540303 (PI: Graham). 
\end{acknowledgements}

\bibliographystyle{aa}
\bibliography{aa57753-25.bib}


\begin{appendix}

\onecolumn

\section{Journal of spectroscopic observations of Nova Vul 2024.}\label{appendix_journal}

We report in the table below a journal of the spectroscopic observations of
NVul24 that we have recorded at the Varese and Asiago observatories.

\begin{table*}[h!]
\caption {Log book of spectroscopic observations of Nova Vul 2024.}
\label{table:3}
\centering
\begin{tabular}{cccccccc}
        &&&&&&\\
	\hline
	&&&&&&\\
	 date       & HJD          & t-t$_{\rm 0}$& expt  & disp.     &  res.pow. & $\lambda$~range    & tel.   \\
                    &              & (days)       & (sec) & (\AA/pix) & $\lambda/\Delta\lambda$ & (\AA) & \\
	&&&&&&\\
	\hline
	&&&&&&\\
	2024-07-30 & 2460522.38 &   0.07 & 2700 &         &  10000  & 4250-8900 & 0.84m \\
	2024-07-31 & 2460523.35 &   1.05 & 2700 &         &  10000  & 4250-8900 & 0.84m \\
	2024-08-01 & 2460524.37 &   2.07 & 2700 &         &  10000  & 4250-8900 & 0.84m \\
	2024-08-02 & 2460525.38 &   3.08 & 1800 &         &  10000  & 4250-8900 & 0.84m \\
	2024-08-03 & 2460526.39 &   4.09 & 2400 &         &  10000  & 4250-8900 & 0.84m \\
	2024-08-05 & 2460528.38 &   6.08 & 2700 &         &  10000  & 4250-8900 & 0.84m \\
    2024-08-06 & 2460529.38 &   7.08 & 2700 &         &  10000  & 4250-8900 & 0.84m \\
	2024-08-07 & 2460530.37 &   8.07 & 2700 &         &  10000  & 4250-8900 & 0.84m \\
	2024-08-09 & 2460532.35 &   10.1 & 1500 &   2.3   &         & 3317-7880 & 1.22m \\
	2024-08-10 & 2460533.33 &   11.0 & 2700 &         &  10000  & 4250-8900 & 0.84m \\
	2024-08-10 & 2460533.36 &   11.1 & 1800 &   2.3   &         & 3317-7880 & 1.22m \\
	2024-08-11 & 2460534.43 &   12.1 & 2400 &   2.3   &         & 3317-7880 & 1.22m \\
	2024-08-12 & 2460535.33 &   13.0 & 2700 &         &  10000  & 4250-8900 & 0.84m \\
	2024-08-13 & 2460536.45 &   14.1 & 2400 &   2.3   &         & 3317-7880 & 1.22m \\
	2024-08-15 & 2460538.37 &   16.0 & 2700 &         &  10000  & 4250-8900 & 0.84m \\
	2024-08-15 & 2460538.46 &   16.1 & 2400 &   2.3   &         & 3317-7880 & 1.22m \\
	2024-08-16 & 2460539.47 &   17.1 & 2400 &   2.3   &         & 3317-7880 & 1.22m \\
	2024-08-17 & 2460540.37 &   18.0 & 2700 &         &  10000  & 4250-8900 & 0.84m \\
	2024-08-22 & 2460545.39 &   23.0 & 2400 &   2.3   &         & 3317-7880 & 1.22m \\
	2024-08-23 & 2460546.32 &   24.0 & 3600 &         &  10000  & 4250-8900 & 0.84m \\
	2024-08-23 & 2460546.38 &   24.1 & 3600 &   2.3   &         & 3317-7880 & 1.22m \\
	2024-08-25 & 2460548.45 &   26.1 & 3600 &   2.3   &         & 3317-7880 & 1.22m \\
	2024-08-26 & 2460549.37 &   27.0 & 2700 &         &  10000  & 4250-8900 & 0.84m \\
	2024-08-30 & 2460553.40 &   31.1 & 3600 &         &  10000  & 4250-8900 & 0.84m \\
	2024-09-02 & 2460556.36 &   34.0 & 3600 &         &  10000  & 4250-8900 & 0.84m \\
	2024-09-06 & 2460560.37 &   38.0 & 3600 &         &  10000  & 4250-8900 & 0.84m \\
	2024-09-10 & 2460564.36 &   42.0 & 3600 &         &  10000  & 4250-8900 & 0.84m \\
	2024-09-21 & 2460575.35 &   53.0 & 1200 &         &  23000  & 3485-7135 & 1.82m \\
	2024-10-29 & 2460575.35 &   91.0 & 3600 &   2.3   &         & 3317-7880 & 1.22m \\
	2024-11-15 & 2460630.21 &  107.9 & 3600 &   2.3   &         & 3317-7880 & 1.22m \\
	2024-11-25 & 2460640.21 &  117.9 & 3600 &   0.6   &         & 5770-7000 & 1.22m \\
	&&&&&&\\
	\hline
	\end{tabular}
	\tablefoot{The interval $t-t_0$ is counted from the epoch of maximum
	optical brightness on 2024 Jul 30.8 UT (HJD=2460522.3). The last column identifies the
        telescopes: Varese 0.84m, Asiago 1.22m, and Asiago 1.82m.}
        \end{table*}

\clearpage
\newpage
\twocolumn

\section{BVRI photometry }\label{appendix_photometry}

Photometry evolution of NVul24 using $B$, $V$, $R$, and $I$ observations
from AAVSO (binned to half a day interval to reduce noise) and 
ZTF $g$ and $r$ data.

 \begin{figure}[ht!]
   \centering
   \includegraphics[width=7cm]{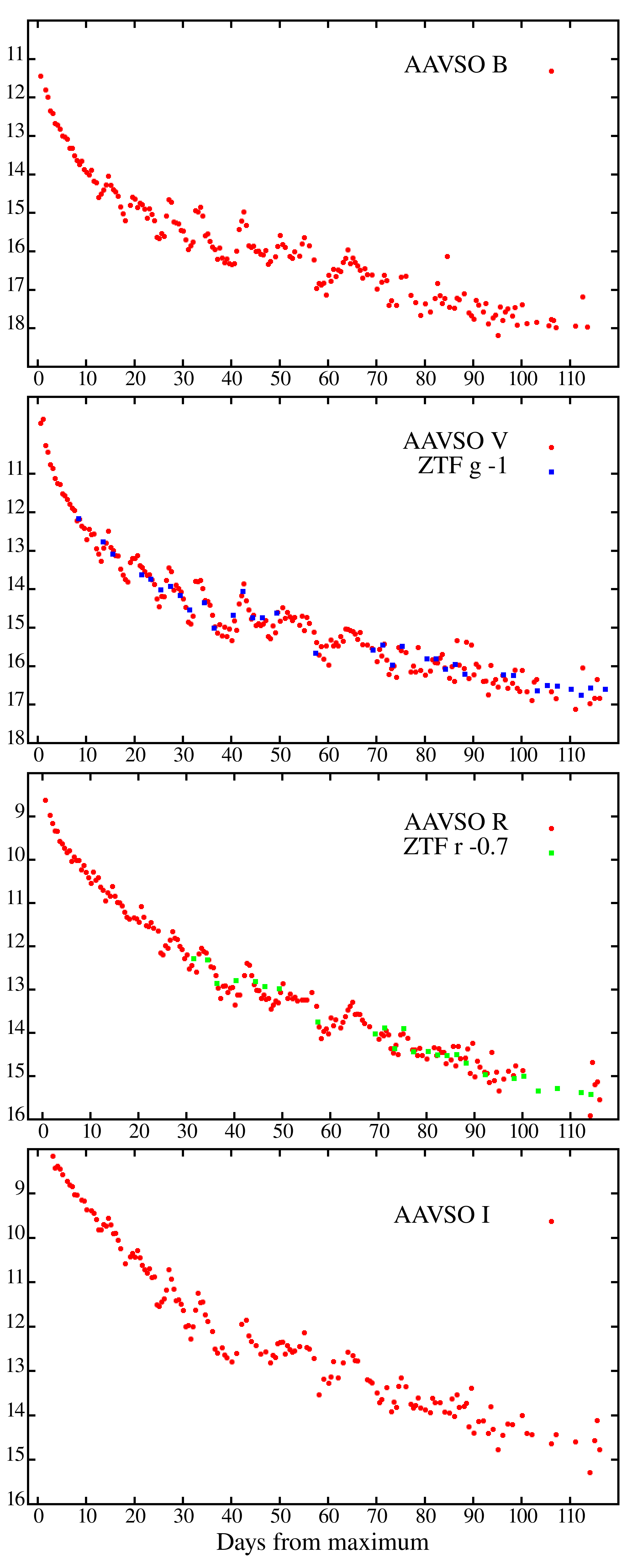}
        \caption{$B$, $V$, $R$, and $I$ band light-curves of Nova Vul 2024.  Red
        dots are from the AAVSO database, binned to half a day to reduce the noise. 
        Blue dots are ZTF $g$ magnitudes shifted to $V$ band by adding a
        $-$1 mag offset, while green dots are ZTF $r$ magnitudes shifted to
        $R$ band by adding a $-$0.7 mag offset.  The earliest three
        measurements in the $V$-band panel, covering the rise to maximum,
        have been imported from CBET 5423.  The observing epochs are counted
        as days passed since optical maximum on 2024 Jul 30.8 UT
        (HJD=2460522.3).}
         \label{phot4}
   \end{figure}

\section{Evolution of Principal and Diffuse Enhanced absorption systems }\label{appendix_abs_systems}

As expected for a very fast nova, the spectral changes of NVul24 in the
first few days after discovery were very rapid, and the “principal” and
“diffuse enhanced” absorptions systems for H$\alpha$
\citep[cf.][]{1960stat.book..585M}, can be followed only for the
first three days past maximum. They are shown in Fig.~\ref{fitting_3} and measured 
Table~\ref{tab_Halfa} below. From day +4 the principal absorption system merged
with the shoulder of the developing boxy profile of H$\alpha$, as visible in
Fig.~\ref{Halpha_OI} .

\begin{figure}[!h]
\centering
   \includegraphics[width=7cm]{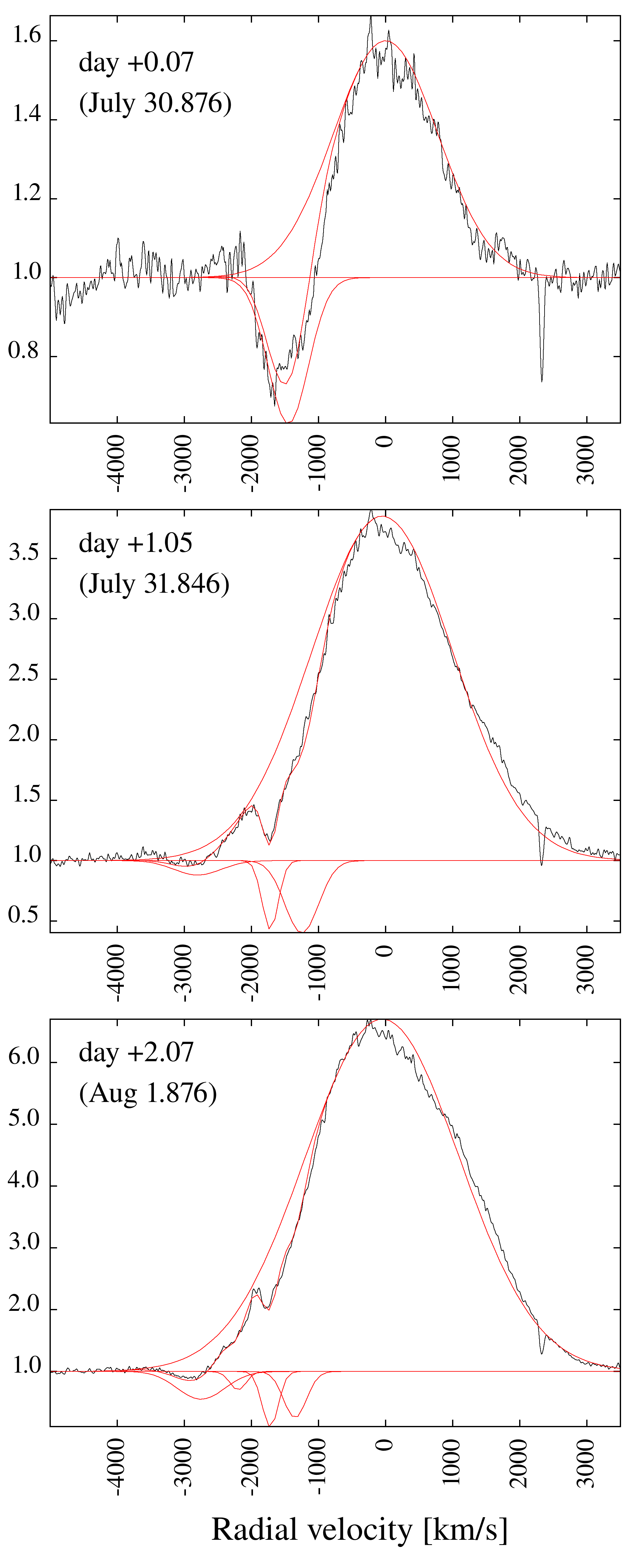}
    \caption{Continuum normalized H$\alpha$ profiles of Nova Vul 2024 for
	the first three days past maximum. The observed spectrum is shown in black, 
    whereas the overall fit and the individual components are shown in red. 
    Their parameters are reported in Table~\ref{tab_Halfa}.}
	\label{fitting_3}
\end{figure}

It is worth noticing that the velocities reported in Table~\ref{tab_Halfa}
well conform with the value predicted by the relations calibrated by 
\citep{1960stat.book..585M} against $t_3$:
\begin{eqnarray}
\log {V_p} &=& 3.70 - 0.5 \cdot \log {t_3} = 1513  ~~{\rm km/s} \\
\log {V_{DE}} &=& 3.81 - 0.4 \cdot \log {t_3} = 2511  ~~{\rm km/s}
\end{eqnarray}
Considering that the (vast) majority of the novae used by \citep{1960stat.book..585M}
to calibrate these relations were of the FeII type (a classification not known
at the time being introduced only much later by
\citet{1992AJ....104..725W}), the agreement with the velocities measured on
our programme nova support the notion that - even if turning hybrid at a later
time - NVul24 was initially behaving like a standard FeII nova, including
the structure and velocity of P-Cyg absorption systems.

   \begin{table}[ht!]
	\centering
	\footnotesize
	\caption{Components of the H$\alpha$ line profile of Nova Vul 2024.}
	\label{tab_Halfa}
	\begin{tabular}{cccccc}
	 &&\\
	\hline\hline
	&&&&\\
	date  &RV$_\odot$           & WHM        & e.w.       & Flux  &  System    \\
	(2024)&      (km/s)         &     (km/s) &      (\AA) &       & \\
	day +0.07                   &0      &  915  & $-$24.7 & 25             & \textit{em.} \\
	(Jul 30.876, UT)            &$-$1450 &  350  &  5.9   & 5.9            & \textit{P}   \\
	&&&&\\
	day +1.05         &   $-$50   &  1235 &$-$158   & 93             & \textit{em.} \\
	(Jul 31.846, UT)  &     $-$1250 &  140  &  3.6   & 2.1            & \textit{P}   \\
	                  & $-$1720 &  290  &  7.9   & 4.6            & \textit{P}   \\
	                  & $-$2800 &  410  &  2.2   & 1.3            & \textit{DE}  \\
	&&&&\\
	 day +2.07 & 	  $-$50   &  1350 & $-$346 & 127            & \textit{em.} \\
	(Aug 1.876,  UT)  &$-$1350 &  200  &  6.7   & 2.5            & \textit{P}   \\
	  &$-$1720 &  140  &  5.7   & 2.1            & \textit{P}   \\
	  &$-$2200 &  140  &  1.9   & 0.7            & \textit{P}   \\
	  &$-$2750 &  410  &  8.3   & 3.0            & \textit{DE}  \\
	&&&&\\
	\hline
	\end{tabular}
        \tablefoot{WHM denotes the width at half maximum; ``em'' is the
	emission component, ``P'' and ``DE'' are the principal and
	diffuse-enhanced absorption systems, respectively.
    The flux is expressed in units of 10$^{-12}$ erg cm$^2$ s$^{-1}$ . }
	\end{table}

\section{Dereddened fluxes of emission lines observed in the spectra
	of NVul24 for days +91 and +107.}

Our latest full-range spectra of NVul24 were obtained on 2024 Oct 29 and Nov
15 (day +91 and +107), and are visible in the top panel of
Fig.~\ref{nebular_ha} after correction for the $E_{B-V}$=1.6 mag reddening. 
They have been recorded when the dilution of the ejecta conformed to
optically thin conditions over the bulk of their mass.  
We report in the table below the integrated flux of the most relevant emission 
lines of NVul24 that we have measured on these spectra.  
For the 4640+HeII 3686 and [OIII] 4959+5007 blends, we assumed the same profile 
for each contributing line, and adjusted the relative fluxes by $\chi^2$ fitting 
to the overall blend profile.  
That is not possible for H$\gamma$+[OIII] 4363 which are almost exactly
superimposed.  

\begin{table}[ht!]
	\centering
	\footnotesize
	\caption{Dereddened fluxes of emission lines observed in the spectra
	of NVul24 for days +91 and +107.}
	\begin{tabular}{lccc}
	\hline\hline
	Name        & Wavelength  & \multicolumn{2}{c}{Flux ($\times$10$^{-12}$ erg cm$^2$ s$^{-1}$)} \\
                    & (\AA)       & day +91 & day +107\\
	\hline
	$[$NeIII$]$            &  3869     &  <10  &  <5     \\
    H$\delta$              &  4102     &  20.0 &  6.0    \\
	$[$OIII$]$ + H$\gamma$ &  4363-40  &  38.0 &  20.0    \\
    blend                  &  4640     &  4.0  &  12.0    \\
    HeII                   &  4686     &  2.0  &  6.0    \\
	H$\beta$               &  4861     &  19.6 &  8.5    \\
	$[$OIII$]$             &  4959     &  10.0 &  8.0    \\
	$[$OIII$]$             &  5007     &  30.0 &  24.0    \\
	$[$FeVI$]$             &  5176     &  2.5  &  1.0    \\
	$[$CaV$]$              &  5309     &  1.0  &  1.2    \\
	$[$NII$]$              &  5755     &  5.7  &  7.5    \\
	HeI                    &  5876     &  3.2  &  1.8    \\
	$[$FeVII$]$+$[$CaV$]$  &  6086     &  3.5  &  2.7    \\
	$[$FeX$]$              &  6375     &  1.3  &  1.6    \\
	H$\alpha$              &  6563     &  59.0 &  49.0    \\
    HeI                    &  7065     &  2.4  &  3.0    \\
	$[$OII$]$              &  7320     &  1.6  &  3.4    \\
		
	\hline
	\end{tabular}
	\label{tab_nebular}
\end{table}

\end{appendix}

\end{document}